\documentclass[12pt,a4paper]{article}
\usepackage{cmap}
\usepackage[T2A]{fontenc}
\usepackage[cp1251]{inputenc}
\usepackage[english, russian]{babel}
\usepackage{amsfonts,amsmath,amssymb,amsthm,longtable,hhline}
\usepackage{bm}
\usepackage{cite}
\usepackage{color,soul}
\usepackage{color,xcolor}
\usepackage{multicol}
\usepackage{tabularx,caption,multirow}
\renewcommand{\arraystretch}{1.2} 

\makeatletter
\renewcommand{\@biblabel}[1]{#1.} 
\makeatother
\theoremstyle{remark}

\newcounter{urav}[section]
\newcounter{resh}[urav]

\newcounter{exmp}

\let\ds=\displaystyle
\let\ts=\textstyle
 
\let\Bl=\Bigl \let\Br=\Bigr
\let\BL=\biggl \let\BR=\biggr
\def\arb{is an arbitrary constant}
\def\arbs{are arbitrary constants}
\def\arbf{is an arbitrary function}
\def\arbfs{are arbitrary functions}
\def\fracskip{\mskip 1mu \relax}
\def\nfrac#1#2{{\fracskip#1\fracskip\over\fracskip#2\fracskip}}
\def\dfrac#1#2{{\ds\nfrac{#1}{#2}}}
\def\tfrac#1#2{{\ts\nfrac{#1}{#2}}}
\let\frac=\nfrac
\def\pd#1#2{\dfrac{\partial#1}{\partial#2}}
\def\pdd#1#2#3{\ifx#2#3\pd{^2#1}{#2^2}\else\pd{^2#1}{#2\partial#3}\fi }

\begin{document}
\large 

\centerline{\bf The direct method of functional separation of variables can provide}
\centerline{\bf more exact solutions than the compatibility analysis of PDEs}
\centerline{\bf based on a single differential constraint}

\bigskip

\centerline{Andrei D. Polyanin$^{a, b, c}$}
\bigskip
\centerline{\it $^a$ Ishlinsky Institute for Problems in Mechanics, Russian Academy of Sciences,}
\centerline{\it 101 Vernadsky Avenue, bldg~1, 119526 Moscow, Russia}
\centerline{\it $^b$ Bauman Moscow State Technical University,}
\centerline{\it 5 Second Baumanskaya Street, 105005 Moscow, Russia}
\centerline{\it $^c$ National Research Nuclear University MEPhI,}
\centerline{\it 31 Kashirskoe Shosse, 115409 Moscow, Russia}
\bigskip
\bigskip

This note shows that in looking for exact solutions to nonlinear PDEs, the direct method of functional separation of variables can, in certain cases, be more
effective than the method of differential constraints based on the compatibility analysis of PDEs with a single constraint (invariant surface condition). This
fact is illustrated by examples of nonlinear reaction-diffusion and convection-diffusion equations with variable coefficients, nonlinear Klein--Gordon type
equations, and hydrodynamic boundary layer equations. A few new exact solutions are given.
\bigskip


\section{Introduction. The methods concerned}\label{s:1}

\subsection{The direct method for constructing functional separable solutions in implicit form}\label{ss:1.1}

Let us look at nonlinear PDEs of the form
\begin{align}
F(x,u_x,u_t,u_{xx},u_{xt},u_{tt},\dots)=0. \label{eq:001}
\end{align}

Equation~\eqref{eq:001} can be analyzed using a direct method of functional
separation of variables based on seeking exact solutions in implicit form
\cite{pol2019c}:
\begin{align}
\int h(u)\,du=\xi(x)\omega(t)+\eta(x). \label{eq:002}
\end{align}
The functions $h(u)$, $\xi(x)$, $\eta(x)$, and $\omega(t)$ are to be determined in a
subsequent analysis.

The procedure for constructing such solutions is as follows. First, using \eqref{eq:002}, one calculates the partial derivatives $u_x$, $u_t$, $u_{xx}$, \dots,
which are expressed in terms of the functions $h$, $\xi$, $\eta$, $\omega$ and their derivatives. Then, these partial derivatives must be substituted into
equation \eqref{eq:001} followed by eliminating the variable $t$ with the help of \eqref{eq:002}. As a result (with a suitable choice of $\omega$), one arrives
at a bilinear functional-differential equation,
\begin{equation}
\begin{gathered}
\sum^N_{j=1}\Phi_j[x]\Psi_j[u]=0.
\label{eq:06**}
\end{gathered}
\end{equation}
Here, $\Phi_j[x]\equiv\Phi_j(x,\xi,\eta,\xi'_x,\eta'_x,\dots)$ and
$\Psi_j[u]\equiv\Psi_j(u,h,h'_u,\dots)$ are differential forms (in some cases,
functional coefficients) that depend, respectively, on $x$ and $u$ alone. The following
statement is true.

\medbreak
 \textit{Proposition} (first formulated by Birkhoff~\cite{bir1960}).
Functional differential equations of the form \eqref{eq:06**} can have solutions only
if the forms $\Psi_j[u]$ ($j=1,\dots,N$) are connected by linear relations (see, for
example, \cite{puc2000,pol2012,pol2019c}):
\begin{align}
\sum^{m_i}_{j=1}k_{ij}\Psi_j[u]=0,\quad \ i=1,\dots,n,
\label{eq:013}
\end{align}
where $k_{ij}$ are some constants, $1\le m_i\le N-1$, and $1\le n\le N-1$.
Degenerate cases must also be treated where, in addition to the linear relations,
some individual differential forms $\Psi_j[u]$ vanish.

\medbreak
This proposition is used for the construction of exact solutions to functional differential
equations of the form \eqref{eq:06**} and the corresponding nonlinear PDEs
\eqref{eq:001}. Note that, in the generic case, different linear relations of the form
\eqref{eq:013} correspond to different solutions of the PDEs under consideration.

\subsection{The method of differential constraints}\label{ss:1.2}

The direct method for constructing functional separable solutions in implicit form based on formula
\eqref{eq:002} is closely related to the method of differential constraints, which is based on the compatibility
theory of PDEs~\cite{yan1964}.

To show this, we differentiate formula \eqref{eq:002} with respect to $t$ to obtain
\begin{align}
u_t=\xi(x)\bar\omega(t)\varphi(u),
\label{eq:003}
\end{align}
where $\bar\omega(t)=\omega^\prime_t(t)$ and $\varphi(u)=1/h(u)$.

Relation \eqref{eq:003} can be treated as a first-order differential constraint, which can be used to find exact
solutions of equation \eqref{eq:001} through a compatibility analysis of the overdetermined pair of equations
\eqref{eq:001} and \eqref{eq:003} with the single unknown~$u$. The differential constraint \eqref{eq:003} is
equivalent to relation \eqref{eq:002}; initially, all functions included on the right-hand sides of
\eqref{eq:002} and \eqref{eq:003} are considered arbitrary, and the specific form of these functions is
determined in the subsequent analysis.

Differential constraints of the second and higher orders can also be used to construct exact solutions to
equation \eqref{eq:001}; in the general case, any PDE (or ODE, in degenerate cases) that depends on the same
variables as the original equation can be treated as a differential constraint. For a description of the method
of differential constraints and its relationship with other methods, as well as a number of specific examples of its
application, see, for example, \cite{yan1964,mel1983,sid1984,gal1994,olv1994,kap1995,and1998,kap2003,pol2012}.
Note that exact solutions can be sought using several differential constraints (see, for example,
\cite{sid1984,pol2012}).

The construction of exact solutions by the method of differential constraints is based on a compatibility
analysis of PDEs and is carried out in several steps briefly described below.

1.\enspace Two PDEs, the original PDE and a differential constraint, are differentiated (sufficiently many times)
with respect to $x$ and $t$, and then the highest-order derivatives are eliminated from the differential
relations obtained and PDEs considered. As a result, one arrives at an equation involving powers of lower-order
derivatives, for example, $u_x$.

2.\enspace By equating the coefficients of all degrees of the derivative $u_x$ with zero in this equation, one
obtains compatibility conditions relating the functional coefficients of the PDEs.

3.\enspace The compatibility conditions make up a nonlinear system of ODEs for determining the functional
coefficients. In this step, it is necessary to find a solution to this system in a closed form.

4.\enspace The obtained coefficients are substituted into the differential constraint, which must then be
integrated to find a form (or forms) of the unknown function $u$ (in this step, intermediate solutions are
obtained that contain undetermined functions).

5.\enspace The final form of the unknown function is determined from the original PDE.

In the last three steps of the method of differential constraints, one has to solve different equations (systems of
equations). If no solution can be found in at least one of these steps, the procedure fails and no
exact solution to the original equation is obtained.
\medskip

\textit{Remark 1.}
The first-order differential constraint \eqref{eq:003} is a special case of an invariant surface condition
\cite{blu1969}, which characterizes the nonclassical method of symmetry reduction. In general, an invariant
surface condition is a quasilinear first-order PDE of general form.
Therefore, the nonclassical method of symmetry reduction can be
considered as an important special case of the method of differential constraints;
specific examples of its use can
be found, for example, in \cite{blu1969,lev1989,nuc1992,cla1995,olv1996,cla1997,puc2000,sac2004,pol2012}.

\subsection{Question: which method is more effective?}\label{ss:1.3}

Although the differential constraint \eqref{eq:003} is equivalent to the functional relation \eqref{eq:002},
the subsequent procedure for finding exact solutions by the direct method for constructing functional separable
solutions in implicit form and that by the method of differential constraints differ significantly. A natural
and very important question arises: Do these methods result in the same exact solutions or not?

It will be shown below that the direct method of functional separation of variables based on the implicit representation of solution \eqref{eq:002}
can provide more exact solutions
than the method of differential constraints
with the equivalent differential constraint (invariant surface condition) \eqref{eq:003}.

\section{Non-linear reaction-diffusion equations with variable\\ coefficients}\label{s:2}

\subsection{Using the method of differential constraints}\label{ss:2.1}

Let us look at nonlinear reaction-diffusion equations with variable coefficients of the form
\begin{align}
c(x)u_{t}=[a(x)f(u)u_x]_x+b(x)g(u). \label{eq:004}
\end{align}

To construct exact solutions to this equation, we use the differential
constraint (invariant surface condition)
\begin{align}
u_t=\theta(x,t)\varphi(u), \label{eq:005}
\end{align}
which is more general than constraint \eqref{eq:003}.

\penalty-50
We solve equation \eqref{eq:004} for the highest derivative and
eliminate $u_t$ with the help of \eqref{eq:005} to obtain
\begin{align}
u_{xx}=-\frac{f'_u}{f}u_x^2-\frac{a'_x}{a}u_x-\frac ba\frac gf+\frac{c\theta}a\frac{\varphi}f. \label{eq:006}
\end{align}

Differentiating \eqref{eq:005} twice with respect to $x$ and taking into account relation \eqref{eq:006}, we get
\begin{align}
u_t&=\theta\varphi,\quad u_{tx}=\theta\varphi'_uu_x+\theta_x\varphi,\notag\\
u_{txx}&=\theta\varphi'_uu_{xx}+\theta\varphi''_{uu}u_x^2+2\theta_x\varphi'_uu_x+\theta_{xx}\varphi\notag\\
&=\theta\Bl(\varphi''_{u}-\frac{f'_u}{f}\varphi'_u\Br)u_x^2+A_1(x,t,u)u_x+A_0(x,t,u). \label{eq:007}
\end{align}
Here $A_1$ and $A_0$ are independent of $u_x$ and are expressed in terms of the functions appearing in PDEs \eqref{eq:004}
and \eqref{eq:005}.

Differentiating \eqref{eq:006} with respect $t$ and taking into account the first two relations of \eqref{eq:007}, we find the
mixed derivative in a different way:
\begin{align}
u_{xxt}=-\theta\Bl[\varphi\Bl(\frac{f'_u}{f}\Br)^{\!\prime}_{\!u}+2\frac{f'_u}{f}\varphi'_u\Br]u_x^2+B_1(x,t,u)u_x+B_0(x,t,u).
\label{eq:008}
\end{align}

By matching up the third-order mixed derivatives \eqref{eq:007} and \eqref{eq:008}, we get the following
relation, quadratic in~$u_x$:
\begin{equation}
\begin{gathered}
F_2u_x^2+F_1u_x+F_0=0,\\
F_2=\theta\Bl[\varphi''_{u}+\varphi'_u\frac{f'_u}{f}+\varphi\Bl(\frac{f'_u}{f}\Br)^{\!\prime}_{\!u}\Br].
\end{gathered}
\label{eq:009}
\end{equation}
The functional coefficients $F_0$ and $F_1$ depend on $a$, $b$, $c$, $f$, $g$, $\theta$, $\varphi$ and
their derivatives (and are independent of $u_x$). By equating the functional coefficients $F_n$ with zero (the
procedure of splitting by the derivative $u_x$), one can obtain a determining system of equations. Next, we
only need the first equation of this system (corresponding to $F_2=0$), which, after dividing by $\theta$, takes
the form
\begin{align}
\varphi''_{u}+\varphi'_u\frac{f'_u}{f}+\varphi\Bl(\frac{f'_u}{f}\Br)^{\!\prime}_{\!u}=0. \label{eq:010}
\end{align}
Considering $f$ to be an arbitrary function and $\varphi$ to be the unknown, we find the general solution of
equation \eqref{eq:010}:
\begin{align}
\varphi=\frac 1f\BL(C_1\int f\,du+C_2\BR),
\label{eq:011}
\end{align}
where $C_1$ and $C_2$ \arbs.
Thus, the method of differential constraints leads to exact solutions in which the functions $f$ and $\varphi$
(involved in the original equation and the differential constraint) are related by \eqref{eq:011}.

Using the differential constraint \eqref{eq:003} is equivalent to
representing the solution in the form \eqref{eq:002}. Since $\varphi=1/h$, solution \eqref{eq:011}
can be rewritten in terms of $f$ and $h$ as
\begin{align}
h=f\BL(C_1\int f\,du+C_2\BR)^{\!-1}.
\label{eq:012}
\end{align}

\subsection{Using the direct method of functional separation of variables}\label{ss:2.2}


The study \cite{pol2019c} presents a large number of exact solutions to PDEs of the form
\eqref{eq:004} obtained using the method described in Section  \ref{ss:1.1}. In particular, it shows that the equation
\begin{align}
u_{t}=[a(x)f(u)u_x]_x+\frac{a'_x(x)}{\sqrt{a(x)}}u, \label{eq:014}
\end{align}
which contains two arbitrary functions $a(x)>0$ and $f(u)$, admits the exact solution in implicit form
\begin{align}
\int\frac{f(u)}u\,du=4t-2\int\frac {dx}{\sqrt{a(x)}}+C, \label{eq:015}
\end{align}
where $C$ \arb.

Solution \eqref{eq:015} is a special case of solutions \eqref{eq:002} with $h=f/u$. This solution
is different from \eqref{eq:012}; consequently, it cannot be obtained by the method of differential constraints
using relation \eqref{eq:003}, neither can it be obtained using the more general differential constraint \eqref{eq:005}.

Solutions of the form \eqref{eq:015} are generated by two differential constraints: one of them is \eqref{eq:003}
and the other (additional) constraint has the form $u_x=p(x)\psi(u)$
(namely, $\sqrt afu_x=-2u$). It is important to note that the latter
constraint is determined by the functional coefficients of the original equation \eqref{eq:004} and cannot be
obtained from general a priori considerations.

In addition to solution \eqref{eq:015}, several other exact solutions of the form \eqref{eq:002} were also obtained
in \cite{pol2019c}, which do not satisfy relation \eqref{eq:012} and are omitted here; just as above, these solutions
cannot be obtained by the method of differential constraints based on a single constraint.
\medskip

\textit{Remark 2.}
It can be shown that solution \eqref{eq:015} cannot be obtained by the method of differential constraints using
a single constraint of the form $u_t=\varphi(x,t,u)$, which is even more general than \eqref{eq:003} and \eqref{eq:005}.

\section{Non-linear convection-diffusion equations with variable coefficients}\label{s:3}

\subsection{Using the method of differential constraints}\label{ss:3.1}

Let us look at nonlinear convection-diffusion equations of the form
\begin{align}
c(x)u_{t}=[a(x)f(u)u_x]_x+b(x)g(u)u_x. \label{eq:004*}
\end{align}

The compatibility analysis of two PDEs, the original equation \eqref{eq:004*} and differential constraint \eqref{eq:005}, is
performed in the same way as in Section~\ref{ss:2.1}. As a result, we obtain a relation, quadratic in $u_x$, in
which the functional coefficient of $u_x^2$ coincides with $F_2$ from \eqref{eq:009}.

Therefore, the method of differential constraints based on the single constraint \eqref{eq:005} for the
convection-diffusion equations \eqref{eq:004*} also results in relations \eqref{eq:011} and
\eqref{eq:012}.

\subsection{Using the direct method of functional separation of variables}\label{ss:3.2}

It can be shown that the nonlinear convection-diffusion equation of special form
\begin{align}
u_{t}=[a(x)f(u)u_x]_x-\tfrac 12 a^\prime_x(x)f(u)u_x, \label{eq:005*}
\end{align}
where $a(x)$ and $f(u)$ \arbfs, admits the pair exact solutions
\begin{align}
\int \frac {f(u)}u\,du=kt\pm \sqrt{k}\int\!\frac {dx}{\sqrt{a(x)}}+C, \label{eq:006*}
\end{align}
with $C$ and $k$ being arbitrary constants.

Solutions \eqref{eq:006*} are special cases of solutions of the form \eqref{eq:002} with $h=f/u$. These solutions
do not satisfy relation \eqref{eq:012} and, therefore, cannot be obtained by the method of differential
constraints based on the single constraint \eqref{eq:003}; however, these solutions can be obtained if two differential
constraints are used at once.

\section{Non-linear Klein--Gordon type equations with variable coefficients}\label{s:4}

\subsection{Using the method of differential constraints}\label{ss:4.1}

Now let us look at the nonlinear Klein--Gordon type equation with variable coefficients
\begin{align}
c(x)u_{tt}=[a(x)f(u)u_x]_x+b(x)g(u). \label{eq:016}
\end{align}

To construct exact solutions to this equation, we also use a more general differential
constraint \eqref{eq:005} than \eqref{eq:003}. Differentiating \eqref{eq:005} with respect to $t$ gives
\begin{align}
u_t=\theta\varphi\quad \Longrightarrow \quad
u_{tt}=\theta\varphi'_uu_t+\theta_t\varphi=\theta^2\varphi\varphi'_u+\theta_t\varphi. \label{eq:017}
\end{align}
We solve equation \eqref{eq:016} for $u_{xx}$ and then eliminate $u_{tt}$ with the help of
\eqref{eq:017} to obtain
\begin{align}
u_{xx}=-\frac{f'_u}{f}u_x^2-\frac{a'_x}{a}u_x-\frac ba\frac gf+\frac
c{af}(\theta^2\varphi\varphi'_u+\theta_t\varphi). \label{eq:018}
\end{align}

Differentiating \eqref{eq:005} with respect to $x$ twice and taking into account relation \eqref{eq:018}, we find
$u_{txx}$. Differentiating \eqref{eq:018} with respect to $t$ and taking into account the first two relations of \eqref{eq:007},
we determine the mixed derivative $u_{xxt}$. By matching up the two third-order mixed derivatives,
$u_{txx}=u_{xxt}$, we arrive at a relation, quadratic in~$u_x$, in which the functional coefficient of~$u_x^2$
coincides with $F_2$ from~\eqref{eq:009}. Using the same reasoning as in Section~\ref{ss:2.1}, we obtain the relation
\eqref{eq:012} between the functions $f$ and $h$ appearing in the equation~\eqref{eq:016} and differential constraint
\eqref{eq:005}.

\subsection{Using the direct method of functional separation of variables}\label{ss:4.2}


Let us look at the nonlinear Klein--Gordon type equation of special form
\begin{align}
u_{tt}=[a(x)f(u)u_x]_x+\frac {x^2}{a(x)}g(u), \label{eq:019}
\end{align}
where $a(x)$ \arbf; the functions $f(u)$ and $g(u)$ are expressed in terms of the arbitrary function $h=h(u)$ as
\begin{align}
f(u)=\frac{h'_u}{h^2},\quad g(u)=-\frac {1}{h}\BL(\frac{h'_u}{h^3}\BR)^{\!\prime}_{\!u}. \label{eq:020}
\end{align}

By the method described in Section \ref{ss:1.1}, we can construct an implicit exact solution to equation
\eqref{eq:019} with $f(u)$ and $g(u)$ defined by \eqref{eq:020}:
\begin{align}
\int h(u)\,du=t-\int\frac{x\,dx}{a(x)}+C. \label{eq:021}
\end{align}

It follows from the first relation of \eqref{eq:020} and solution \eqref{eq:021} that relation \eqref{eq:012} is not
satisfied, and hence, solution \eqref{eq:021} cannot be obtained by the method of differential constraints
with the single constrain \eqref{eq:003}.

\section{Axisymmetric boundary layer equations}\label{s:5}

\subsection{Functional separable solutions in explicit form}\label{ss:5.1}

The system of equations of a laminar unsteady axisymmetric boundary layer on a body of revolution can be reduced through the introduction of a stream function $w$
(and a suitable new independent variable $z$) to a single nonlinear third-order PDE with variable
coefficients~\cite{polzhu2015b}:
\begin{align}
w_{tz}+w_zw_{xz}-w_xw_{zz}=\nu r^2(x)w_{zzz}+F(t,x), \label{eq:03}
\end{align}
where $r=r(x)$ describes the shape of the body (this function is considered arbitrary
here), while $F(t,x)$ defines the pressure gradient.

Exact solutions to equation \eqref{eq:03} can be sought using the method of functional separation of variables in the
explicit form~\cite{polzhu2015b}
\begin{align}
w=fu(\xi)+gz+h,\quad \ \xi=\varphi z+\psi, \label{eq:001*}
\end{align}
with the functions $f=f(t,x)$, $g=g(t,x)$, $h=h(t,x)$, $\varphi=\varphi(t,x)$, $\psi=\psi(t,x)$, and $u=u(\xi)$
to be determined. Substituting \eqref{eq:001*} into equation \eqref{eq:03} and replacing $z$ with
$(\xi-\psi)/\varphi$ yields the functional differential equation
\begin{align}
\sum^6_{n=1}\Phi_n[t,x]\Psi_n[\xi]=\Psi_7[\xi]. \label{eq:002*}
\end{align}
Here, $\Phi_n[t,x]$ are differential forms dependent on the functional coefficients (and their derivatives) involved in \eqref{eq:001*} and \eqref{eq:03},
with all $\Phi_n$ being independent of~$u$.
The forms $\Psi_n=\Psi_n[\xi]$ are expressed as~\cite{polzhu2015b}
\begin{equation}
\begin{aligned}
\Psi_1&=1,\quad \Psi_2=u'_\xi,\quad \Psi_3=(u'_\xi)^2,\quad \Psi_4=u''_{\xi\xi},\\
\Psi_5&=\xi u''_{\xi\xi},\quad \Psi_6=uu''_{\xi\xi},\quad \Psi_7=u'''_{\xi\xi\xi}.
\end{aligned}
\label{eq:100}
\end{equation}

The variables in equation \eqref{eq:002*} can be separated if we assume that the $\Phi_n[t,x]$ on the left-hand
side of \eqref{eq:002*} are all proportional to $r^2f\varphi^3$. This leads to an overdetermined system of PDEs,
\begin{align}
\Phi_n[t,x]=a_n,\quad \ n=1,\dots,6\quad \ (a_n=\text{const}), \label{eq:002*}
\end{align}
and a nonlinear ODE for $u=u(\xi)$,
\begin{equation}
\sum_{n=1}^6a_n\Psi_n=\Psi_7. \label{eq:101}
\end{equation}

If, for some $a_n$, one succeeds in finding a particular solution to the nonlinear system \eqref{eq:002*}, then the
corresponding solution to equation \eqref{eq:101} will generate an exact solution to equation \eqref{eq:03}.

\subsection{Using multiple differential constraints}\label{ss:5.2}

It can be shown that the most interesting solutions of the form \eqref{eq:001*}, those involving several arbitrary
functions, may be obtained if one uses two or three differential relations that are linear combinations
of the forms $\Psi_n$ defined in \eqref{eq:100}.

Table \ref{tab:1} lists a number of functions $u=u(\xi)$ that generate two or three linear differential constraints among the
differential forms \eqref{eq:100}. The differential constraints shown in the first ten rows were described in
\cite{polzhu2015b}; the last four rows show new differential constraints, which generate new exact
solutions of the form \eqref{eq:001*} to equation \eqref{eq:03}.

\begin{table}[!ht]
\small
\renewcommand{\arraystretch}{1.2}
\setlength{\tabcolsep}{4pt}
\thickmuskip=2mu
\medmuskip=1mu
\begin{center}
\caption{Generating functions $u$ and the corresponding linear relations among $\Psi_n$.}
\vspace{1ex}\small
\begin{tabular}{|c|c|c|}
\hline
 No.& Generating functions $u$& Linear constraints between $\Psi_n$\\ \hline
 1& $u=\xi^2$& $\Psi_4=2\Psi_1$, \ $\Psi_5=\Psi_2$, \ $\Psi_6=\tfrac12\Psi_3$ \\ \hline
 2& $u=\xi^3$& $\Psi_5=2\Psi_2$, \ $\Psi_6=\tfrac23\Psi_3$, \ $\Psi_7=6\Psi_1$ \\ \hline
 3& $u=\xi^4$& $\Psi_5=3\Psi_2$, \ $\Psi_6=\tfrac34\Psi_3$ \\ \hline
 4& $u=\xi^{-1}$& $\Psi_5=-2\Psi_2$, \ $\Psi_6=2\Psi_3$, \ $\Psi_7=-6\Psi_3$ \\ \hline
 5& $u=\xi^n$& $\Psi_5=(n-1)\Psi_2$, \ $\Psi_6=\tfrac{n-1}n\Psi_3$\ \ $(n\ne-1,0,1,2,3)$ \\ \hline
 6& $u=\exp \xi$& $\Psi_2=\Psi_4=\Psi_7$, \ $\Psi_6=\Psi_3$ \\ \hline
 7& $u=\cosh \xi$& $\Psi_6=\Psi_1+\Psi_3$, \ $\Psi_7=\Psi_2$ \\ \hline
 8& $u=\sinh \xi$& $\Psi_6=\Psi_3-\Psi_1$, \ $\Psi_7=\Psi_2$ \\ \hline
 9& $u=\cos \xi$& $\Psi_6=\Psi_3-\Psi_1$, \ $\Psi_7=-\Psi_2$ \\ \hline
 10& $u=\sin \xi$& $\Psi_6=\Psi_3-\Psi_1$, \ $\Psi_7=-\Psi_2$ \\ \hline
 11& $u=\tanh \xi$& $\Psi_6=-2\Psi_2+2\Psi_3$, \ $\Psi_7=-2\Psi_2-3\Psi_6$ \\ \hline
 12& $u=\coth \xi$& $\Psi_6=-2\Psi_2+2\Psi_3$, \ $\Psi_7=-2\Psi_2-3\Psi_6$ \\ \hline
 13& $u=\tan \xi$& $\Psi_6=-2\Psi_2+2\Psi_3$, \ $\Psi_7=2\Psi_2+3\Psi_6$ \\ \hline
 14& $u=\cot \xi$& $\Psi_6=2\Psi_2+2\Psi_3$, \ $\Psi_7=2\Psi_2-3\Psi_6$ \\ \hline
\end{tabular}
\label{tab:1}
\end{center}
\end{table}

It is important that the differential constraints specified in Table \ref{tab:1}
are not known in advance. They arise in the course of the analysis and
result from the representation of solutions to equation \eqref{eq:03} in the
form of \eqref{eq:001*} and while using equation \eqref{eq:101}.

Similar exact solutions based on several differential connections for other hydrodynamic boundary layer equations
are obtained in \cite{polzhu2016,polzhu2016a}.

\section{A note on the direct method of Clarkson and Kruskal}

Let us now briefly discuss the direct method of Clarkson and Kruskal \cite{cla1989} (see also \cite{olv1994,cla1997,lud2000,sac2004,pol2012,lud1999}), which is
based on looking for exact solutions in the form $u=U(x,t,w(z))$ with $z=z(x,t)$. The functions $U(x,t,w)$ and $z(x,t)$ should be chosen so as to obtain
ultimately a single ordinary differential equation for $w=w(z)$. The requirement that the function $w$ must satisfy a single ODE greatly limits the
capabilities of this method and does not allow it to be effectively used to find exact solutions such as presented in this note.

The effectiveness of the direct method of Clarkson and Kruskal will increase significantly
if we assume that the function $w$ can satisfy an overdetermined system of several ODEs (see, for example, Section \ref{ss:5.2}).

\section*{Acknowledgments}\addcontentsline{toc}{section}{Acknowledgments}

The study was supported within the framework of the Russian State Assignment
(State Registration Number AAAA-A17-117021310385-6) and partially supported by
the Russian Foundation for Basic Research (project No. 18-29-10025).

I would like to express my deep gratitude to Alexei Zhurov and Alexander Aksenov for fruitful discussions.

\renewcommand{\refname}{References}

\end{document}